\documentstyle[12pt]{article}
\begin{document}
\thispagestyle{empty}
\begin{center}
\LARGE \tt \bf {Non-Riemannian acoustic spacetime of vortex hydrodynamics in Bose-Einstein condensates}
\end{center}
\vspace{5cm}
\begin{center}
{\large L.C. Garcia de Andrade\footnote{Departamento de F\'{\i}sica Te\'{o}rica - Instituto de F\'{\i}sica - UERJ 
Rua S\~{a}o Fco. Xavier 524, Rio de Janeiro, RJ
Maracan\~{a}, CEP:20550-003.E-mail: garcia@dft.if.uerj.br}}
\end{center}
\vspace{2cm}
\begin{abstract} 
Applications of non-Riemannian acoustic geometries in Bose-Einstein condensates (BEC) are considered. The first is the Minkowski-Cartan irrotational vortex acoustic geometry of nonlinear Schr\"{o}dinger equations of BEC (Gross-Pitaeviskii (GP) equation). In this model, which is an alternative to the Riemannian acoustic geometry of phonons in BEC, the Cartan acoustic torsion is physically interpreted as the bending of the BEC wave function amplitude. Actually this shows that acoustic torsion is given by the density perturbation of BEC flow as happens in relativistic cosmological fluid spacetimes. The Ricci-Cartan curvature scalar is computed and a torsion singularity is found at the origin of a quantized vortex in BEC. In the second example, a transverse Magnus force is shown to be expressed in terms of acoustic torsion on a teleparallel vortex acoustics geometry. 
\end{abstract}

\newpage
\section{Introduction}
Recently we have proposed \cite{1} of non-Riemannian geometry of vortex acoustics in classical hydrodynamics, where an example has been given of the equivalence between Letelier's \cite{2} teleparallel torsion loop metric with the superfluid $^{4}He$ metrics \cite{3}. In this work we proposed two new examples of the methods of non-Riemannian effective geometry. The first represents the Minkowski-Cartan framework of the GP BEC equation, where the GP equation is shown to be expressed as the analog of the relativistic particle wave equation on this non-Riemannian spacetime with nonlinear sources. This model serve as an alternative to the Riemannian acoustic analog gravity developed by Unruh, Visser and collaborators \cite{4,3,5}. It is important to stress that as happens with teleparallel gravity is not always worth to work with non-Riemannian BEC analog gravity models instead that their Riemannian counterparts, however, in some examples as the problem of the effective geometrization of the vorticity in fluids, discovered by Bergliaffa et al \cite{6} this alternative of introducing the non-Riemannian \cite{1} relativistic wave equation in analog gravity framework leads to simple explanation of the vorticity of superfluids in terms of the Cartan acoustic torsion. By the way, in our second example which deals with the teleparallel acoustic geometry of moving vortex in hydrodynamical BEC \cite{7}, one shows that the Cartan acoustic torsion which is again proportional to the vorticity of the fluid. In the first case the alternative Riemannian BECs , would also represent a mathematical simpler model , since in Minkowski-Cartan case, only the Cartan axial-vector survives the metric being triavially the Minkowski flat metric. The Minkowski-Cartan relativistic wave equation is on the space of phases, the velocity of fluid being obtained through the gradient of the phase of the BEC wave function. The second example on a certain extent ,although possesses acoustic torsion is equivalent to the Riemannian approach to BEC, since teleparallelism is equivalent to the general theory of relativity which is built on a Riemannian background. The paper is organised as follows: In section 2 we present a very brief review of the Riemannian acoustic geometry of BEC. Section 3 deals with the teleparallel acoustic vortex geometry of the transverse vortex motion throughout the fluid where the Magnus force is simply shown to be proportional to the Cartan acoustic torsion. Section 4 contains the teleparallel acoustic  geometry of BEC. Section 5 presents the conclusions and discussions.  
\section{Riemannian acoustic geometry in BEC}
The effective gravity arises when the effective acoustic metric of BEC \cite{8} is given in the pseudo-Riemannian Lorentzian effective geometric approach. BECs are described mathematically by the nonlinear Schr\"{o}dinger equation:
\begin{equation}
i\hbar\frac{{\partial}{{\Psi}(\vec{r},t)}}{{\partial}t}=[-\frac{{\hbar}^{2}}{2m}{\nabla}^{2}+ V_{ext}(\vec{r})+{\lambda}|{\Psi}|^{2}]{\Psi}
\label{1}
\end{equation}
Using the Madelung transformation
\begin{equation}
{\Psi}(\vec{r},t)= \sqrt{\rho} exp(-i\frac{{\theta}m}{\hbar})
\label{2}
\end{equation}
one obtains after substitution into equation (\ref{1}) the following hydrodynamic form for the GP equation
\begin{equation}
\frac{{\partial}{\theta}}{{\partial}t}+\frac{1}{2}({\nabla}{\theta})^{2}+ \frac{{\lambda}{\rho}}{m}-\frac{{\hbar}^{2}}{2m^{2}}\frac{{\nabla}^{2}\sqrt{\rho}}{\sqrt{\rho}}=0
\label{3}
\end{equation}
the other equation is the continuity equation
\begin{equation}
{\partial}_{t}{\theta}+{\nabla}.({\rho}{\nabla}{\theta})=0
\label{4}
\end{equation}
where the flow velocity is given in the irrotational form by $\vec{v}:={\nabla}{\theta}$. The speed of sound is
\begin{equation}
{c_{s}}^{2}=\frac{dp}{d{\rho}}={\lambda}\frac{\rho}{m}
\label{5}
\end{equation}
Phonons propagating in this effective spacetime sees the metric $g_{ij}$ in the form of a Painleve-Gullstrand form 
\begin{equation}
g_{00}=-\frac{{\rho}}{c}[c^{2}-v^{2}],  g_{0a}=-\frac{\rho}{c}v_{a}
\label{6}
\end{equation}
\begin{equation}
g_{ab}=\frac{\rho}{c}{\delta}_{ab}
\label{7}
\end{equation}
where the indices are $(i,j=0,1,2,3)$ and $(a,b=1,2,3)$. Here c represents the sound speed of phonons. In the next section we shall show that the GP equation in hydrodynamical form can be written as an acoustic Minkowski-Cartan spacetime where the phonons sees the Minkowski acoustic spacetime with an acoustic torsion perturbation caused by the inhomogeneity in BEC. 
\section{Non-Riemannian acoustic BEC spacetime}
Despite of the mathematical ellegance of the Riemannian acoustic geometry , either relativistic \cite{9} or non-relativistic, they in general represent a metric with four metric components which sometimes lead to long computation of curvature and other differential geometric machinery. One of the advantage of the framework we shall now describe isthat we do not work with a metric tensor directly in the computations but only with the acoustic torsion vector part. This , to start with makes the computations much simpler and the physical consequences are similar to the ones obtained by using the Riemannian BEC. In the analog gravity spirit we in general choose a relativistic theory of gravity as a patch to follow. In our case here we consider a gravitational theory in Minkowski spacetime endowed with Cartan torsion to describe the analog system in BEC. This spacetime shall be called an acoustic spacetime where Cartan torsion shall be given by the inhomogeneity in the BEC vapour. A little but straightforward algebra on the GP-Madelung system leads to the equation 
\begin{equation}
{\Box}{\theta}+ \vec{K}.{\nabla}{\theta}=2{\partial}_{t}({\rho}{\epsilon})
\label{8}
\end{equation}
where the Ricci-Cartan scalar curvature is given by
\begin{equation}
R(K)= {\nabla}.\vec{K}+6{\vec{K}}^{2}
\label{9}
\end{equation}
and $\vec{K}=\frac{{\nabla}{\rho}}{\rho}$ is the spatial vector given by the contration of contortion tensor ${K^{i}}_{ji}$ while $K^{0}$ component vanishes. The physical interpretation for the acoustic torsion can thus be given in terms of the cosmological-like density perturbation. This expression for torsion is also suitable to be interpreted as the surface gravity. Unfortunatly here, since the basic acoustic metric is Minkowskian , no acoustic spacetime horizon exists. To simplify matters we consider just the linear approximation on $\vec{K}$, ${\Box}$ is the Minkowskian $M_{4}$ D'Alembertian wave operator and ${\epsilon}$ is the energy density given by
\begin{equation}
{\epsilon}= \frac{1}{2}{\rho}{\vec{v}}^{2}+ {{V}_{eff}}^{(BEC)}
\label{10}
\end{equation}
where ${{V}_{eff}}^{(BEC)}$ is the interaction energy of BEC in non-Riemannian the acoustic spacetime.
\noindent
\begin{equation}
\label{11}
\end{equation}
\begin{equation}
{{V}_{eff}}^{(BEC)}= \frac{{\lambda}^{-1}}{2}\frac{{\hbar}^{2}{c_{s}}^{2}R(K)}{{\rho}\sqrt{\rho}}
\label{12}
\end{equation}
Thus we transform the GP equation of BEC to a non-Riemannian wave equation ,for the phase of the order parameter, with nonlinear sources in acoustic spacetime. The Ricci scalar curvature can be also expressed as 
\begin{equation}
R(K)= {\lambda}\frac{{\nabla}^{2}{\rho}}{\rho}+ \frac{(6{\lambda}-1)}{{\lambda}}{\vec{K}}^{2}
\label{13}
\end{equation}
Let us now compute the Ricci scalar in the non-Riemanian acoustic spacetime of a quantized vortex in the fluid. This can be obtained by the Schr\"{o}dinger equation for this physical system which is given by
\begin{equation}
-\frac{{\hbar}^{2}}{2m^{2}}\frac{{\nabla}^{2}{\sqrt{\rho}}}{\sqrt{\rho}}+ \frac{{\hbar}^{2}}{2m^{2}r^{2}}+\frac{V_{eff}}{m}=0
\label{14}
\end{equation}
from this equation it is easy to compute the Ricci scalar $R(K)$ in non-Riemannian acoustic spacetime, as
\begin{equation}
R(K)= {\lambda}\frac{{\nabla}^{2}{\sqrt{\rho}}}{\sqrt{\rho}}=\frac{\lambda}{r^{2}}+
\frac{mV_{eff}}{{\hbar}^{2}}
\label{15}
\end{equation}
Note that going toward the center of the vortex at $r=0$ the expression (\ref{15}) reduces to
\begin{equation}
R^{vortex}(K)= \frac{\lambda}{r^{2}}
\label{16}
\end{equation}
Note that this expression presents an acoustic torsion singularity at the quantized vortex centre. It also represent that the phonons autoparallels deviate from the initially parallel lines of trajectory towards the vortex core. Actually by geodesic deviation equation with torsion ,it is easy to show that acoustic torsion acts locally as a convergence lens in acoustic spacetimes to which concerns the trajectory of phonon rays. Yet we observe that at very far distances from the vortex core the Ricci-Cartan scalar becomes
\begin{equation}
R(K)= \frac{mV_{eff}}{{\hbar}^{2}}
\label{17}
\end{equation}
It is also distinct to the expression obtained by Fischer and Visser \cite{10} for the Ricci scalar in Riemannian acoustic spacetime, which is given by
\begin{equation}
R^{vortex}({\Gamma})= \frac{{\epsilon}_{0}}{r^{4}}
\label{18}
\end{equation}
where ${\Gamma}$ is the Riemannian-Christoffel connection and ${\epsilon}_{0}$ is given by ${\epsilon}_{0}=\frac{{\Gamma}^{*}}{\sqrt{2}c}$ where ${\Gamma}^{*}=\frac{h}{m}$ is the quantum circulation of the quantized vortex. The non-Riemannian acoustic spacetime Ricci scalar can be also expressed in terms of the quantum circulation as 
\begin{equation}
R(K)= \frac{V_{eff}}{h{\Gamma}^{*}}
\label{19}
\end{equation}
Here ${\epsilon}_{0}$ has the order of the atomic size. As in reference \cite{} experiments based on the focusing of phonon rays around the vortex could be proposed to check the non-Riemannian acoustic spacetime in dilute BEC. 
\section{Magnus force in teleparallel acoustic spacetime in BEC}
In this section we shall give another interesting example of non-Riemannian acoustic spacetime in BEC, namely the teleparallel \cite{11} acoustic geometry in acoustic spacetime. This geometry has the advantage of using the already known Riemannian acoustic metrics of BECs since in this case the Cartan acoustic torsion can be expressed directly in terms of the metric by the expression
\begin{equation}
S_{ijk}= g_[jk,i]-g_[ik,j]
\label{20}
\end{equation}
where the comma represents partial derivatives while the square brakets represents anti-symmetrization. The metric $g_{ij}$ is the Riemannian acoustic spacetime metric. Accordingly to the BEC metric above one may compute the connections and torsion components as
\begin{equation}
{\Gamma}_{ab0}= \frac{\rho}{c}({\partial}_{b}v_{a}-{\partial}_{a}v_{b})
\label{21}
\end{equation}
which yields 
\begin{equation}
{\Gamma}_{ab0}= \frac{\rho}{c}{\Omega}_{ba}
\label{22}
\end{equation}
where ${\Omega}_{ab}$ is the vorticity tensor of the fluid. This result is agrees with the non-Riemannian acoustic torsion model in reference \cite{} where the rotation of the superfluid $^{4}He$ has been associated with acoustic torsion. The other nonvanishing torsion component is
\begin{equation}
S_{0b0}= \frac{\rho}{c}[{\partial}_{b}(c^{2}-v^{2})-{\partial}_{0}v_{b}]
\label{23}
\end{equation}
As shown by P.M. Zhang et al \cite{8} the vortex moving in the fluid shall be exerted a transverse force force which is just the Magnus force. In the light of the teleparallel \cite{11} the acoustic geometry Zhang et al expression may be rewritten as 
\begin{equation}
F_{tr}= {\rho}S_{120}[v^{vortex}-v_{n}]
\label{24}
\end{equation}
This shows that torsion may indeed contribute to Magnus force. Actually a more general acoustic torsion contribution to Magnus force could be obtained from the equation of motion on a torsioned background
\begin{equation}
\frac{d^{2}{x^{i}}}{ds^{2}}= -[{{\Gamma}^{i}}_{jk}-{S^{i}}_{(jk)}]\frac{dx^{j}}{ds}\frac{dx^{k}}{ds}
\label{25}
\end{equation}
Unfortunatly in the first example the torsion contribution to the Magnus force is null since the acoustic torsion there is only due to the vector part of torsion, so an experimental proposal to this non-Riemannian acoustic spacetime geometry of BEC could not be done along the lines of the Magnus force as long as this model is concerned, but it certainly can by making use of the focusing of phonon rays around a torsion string vortex.
\section{Conclusions}
Two physical examples of non-Riemannian acoustic spacetimes. The first is the interpretation of the GP equation on an $M_{4}$ acoustic spacetime with torsion, where a torsion singularity in the scalar curvature is found. Acoustic torsion is shown to be interpreted as a surface gravity or a density perturbation on a BEC cosmological setup. The second example concerns the interpretation of the Riemannian acoustic spacetime of BEC written in terms of a teleparallel acoustic geometry. In this case the rotation of the fluid equals one of the components of acoustic torsion. Magnus force is also shown to be expressed in terms of torsion not only in teleparallelism but in any non-Riemannian acoustic geometry where acoustic torsion is not totally skew symmetric or the trace of the torsion tensor. Experiments can be suggested to investigate these models within the basis of focusing of the phonon rays around a vortex string in BEC.

\end{document}